%% file: main.tex
\documentclass[manuscript,balance=false]{acmart} 
\usepackage{tabularx}
\usepackage{subcaption}
\usepackage{multirow}
\usepackage{framed}
\usepackage{xspace}
\usepackage{pifont}
\usepackage[normalem]{ulem}

\copyrightyear{YYYY}

\acmYear{YYYY}
\acmVolume{YYYY}
\acmNumber{X}
\acmDOI{XXXXXXX.XXXXXXX}
\acmISBN{}
\acmConference{}
\begin{document}

\title{Lost in Permissions: Exploring the Microsoft 365 App Ecosystem}


\author{Vincenzo Longo}
\affiliation{%
  \institution{Politecnico di Torino}
  \city{Torino}
  \country{Italy}}
\email{vincenzo.longo@studenti.polito.it}

\author{Alberto Verna}
\affiliation{%
  \institution{Politecnico di Torino}
  \city{Torino}
  \country{Italy}}
\email{alberto.verna@polito.it}

\author{Nikhil Jha}
\affiliation{%
  \institution{Politecnico di Torino}
  \city{Torino}
  \country{Italy}
}
\email{nikhil.jha@polito.it}

\author{Marco Mellia}
\affiliation{%
 \institution{Politecnico di Torino}
 \city{Torino}
 \country{Italy}}
\email{marco.mellia@polito.it}


\renewcommand{\shortauthors}{Longo\ et al.}

\begin{abstract}
The Microsoft 365 (M365) ecosystem hosts thousands of third-party applications that integrate with enterprise tenants via fine-grained OAuth permissions, potentially granting access to sensitive organisational resources such as emails, files, calendars, chats, and user directories. Despite the security implications of these permission grants, the M365 ecosystem has not been systematically studied.

We present the first privacy- and security-oriented measurement of M365 third-party applications. By combining public marketplace APIs with automated tenant-side deployment, we crawl over 8,000 applications. {We find that only 1,069 of them expose both descriptions and permission sets, with significant inconsistencies in transparency across official distribution channels.}

Next, we leverage a topic-aware anomaly detection framework to assess whether requested permissions align with declared functionality. We cluster applications via Neural Topic Modelling and apply unsupervised anomaly detection within each topic to identify deviations from peer permission profiles. LLM-assisted analysis of the most anomalous cases and blind manual inspection reveal a correlation between anomalous permission profiles and the risk associated with the requested permissions. We find that many applications request overly broad tenant-wide scopes (e.g., directory-wide read/write access), violating least-privilege principles and increasing the organisational attack surface. Our pipeline provides tenant administrators with actionable insights by identifying anomalous applications and the permissions that most contribute to their anomalousness.

Overall, our findings expose systemic opacity and structural immaturity in the M365 app ecosystem, where permission disclosure is inconsistent and over-privileged access is common.


\end{abstract}


\maketitle

\input{text/1-introduction}
\input{text/2-m365}

\input{text/3-dataset}
\input{text/4-clustering}
\input{text/5-anomaly-detection}
\input{text/6-validation}
\input{text/7-related-work}
\input{text/8-conclusion}

\begin{acks}
This work was supported by the AI4CTI FISA under Project \#FISA-2023-00168, funded by the Italian Ministry of University and Research (MUR).

The authors used generative AI-based tools to revise the text, improve flow and correct any typos, grammatical errors, and awkward phrasing. The authors are fully responsible for the content of the paper, including the study design, data collection, analysis, interpretation of results, and all conclusions.
\end{acks}

\bibliographystyle{ACM-Reference-Format}
\bibliography{bibliography}

\appendix
\input{text/998-appendix-short}

\end{document}

%% file: text/1-introduction.tex
\section{Introduction}

Microsoft 365 (M365) is widely deployed in enterprise environments and enables third-party applications to integrate with tenants through a fine-grained OAuth permission model, which comprises more than 300 distinct permissions.
These may grant access to highly sensitive organisational resources, including emails, files, chats, calendars, and tenant-wide user directories. While applications extend platform functionality, the permissions they request may also introduce significant privacy and security risks when they request excessive or misaligned access. %

The M365 ecosystem offers multiple channels to install apps: the official Teams and Office add-ins stores, the global M365 Marketplace, and, lastly, any generic third-party app can access M365 functionalities via API. 
Curiously, permission disclosure is inconsistent across distribution channels: some stores expose detailed permission lists; others omit them; some do not even require explicit consent during installation.

Security best practice requires applications to follow the principle of least privilege, requesting only permissions strictly necessary for their functionality.
This may not always be the case, with developers who can ask for permissions that are unrelated or beyond the application's true needs, causing potential privacy issues or malicious access to the user's and organisation's resources. Unlike mobile ecosystems~\cite{yang2025guidelines, g-cata_2024, BERTDetect_2025,mohd2024ios, scoccia2022empirical}, M365 apps can obtain tenant-wide scopes, exposing organisation-wide data and increasing the attack surface of enterprise environments.

Despite the scale and sensitivity of the M365 ecosystem, no prior work has systematically analysed the permission landscape of third-party applications or assessed whether requested scopes align with declared functionality.
In this work, we compile the first dataset of M365 applications by aggregating data from multiple sources: Marketplace APIs, store Web scraping, and applications installed by our University’s M365 service users. Out of more than 8,000 apps we crawled, only 1,069 expose both a textual description and the list of requested permissions -- the remaining ones simply do not make this information publicly accessible. Our dataset provides the first consolidated view of permission usage in the M365 app ecosystem.

Next, we propose a topic-aware anomaly detection framework that models permission profiles within semantically coherent clusters of applications, enabling detection of deviations relative to functional peers rather than across unrelated apps. We apply Neural Topic Modelling (NTM) to cluster applications according to their declared functionality, obtaining 24 semantic topics.
To identify applications whose permission sets deviate from those of their peers, we apply unsupervised anomaly detection (AD) within each topic.
We highlight both over- and under-privileged applications, which we validate through LLM-assisted evaluation and blind manual inspection. We identify applications requesting tenant-wide read/write access to directories, calendars, chats, and files without clear alignment to their declared purpose, as well as applications lacking permissions required to accomplish their stated goals. {Results indicate that topic-aware anomaly detection captures meaningful deviations in permission usage that are often associated with potentially risky permissions. As such, anomaly scores provide tenant administrators with an additional criterion for assessing whether an application deserves further scrutiny before approval.}
    
In summary, this paper makes {four} contributions:
\begin{itemize}
        \item we construct the first consolidated dataset of M365 third-party applications, combining marketplace APIs, automated deployment, and operational tenant data;
        \item we introduce a topic-aware anomaly detection framework to assess permission-functionality alignment;
        \item we empirically characterise systemic inconsistencies and over-privileged access patterns in the M365 ecosystem;
        \item we present our pipeline as a decision-support tool for evaluating the risk of granting an application access to a tenant.
\end{itemize}
To allow reproducibility and further investigation, we share both the collected dataset and the analysis pipeline.\footnote{Code and dataset will be shared upon the paper's acceptance, to preserve authors' anonymity.}

The rest of the paper is structured as follows. In Section~\ref{sec:m365}, we offer a brief overview of the Microsoft 365 ecosystem. In Section~\ref{sec:dataset} and Section ~\ref{sec:clustering}, we describe the process of retrieving the dataset and running a semantically-coherent clustering of the applications, respectively. In Section~\ref{sec:anomaly-detection} we show the outcome of the anomaly detection algorithms, and in Section~\ref{sec:validaton} we observe the correlation between an app's anomaly and its riskiness. We summarise related works in Section~\ref{sec:related} before drawing conclusions and outlining perspectives for future research on the topic in Section~\ref{sec:conclusion}.

%% file: text/2-m365.tex
\section{The Microsoft 365 ecosystem}
\label{sec:m365}

\begin{figure*}[t]
  \centering
  \begin{subfigure}[b]{0.28\textwidth}
    \centering
    \includegraphics[width=\textwidth]{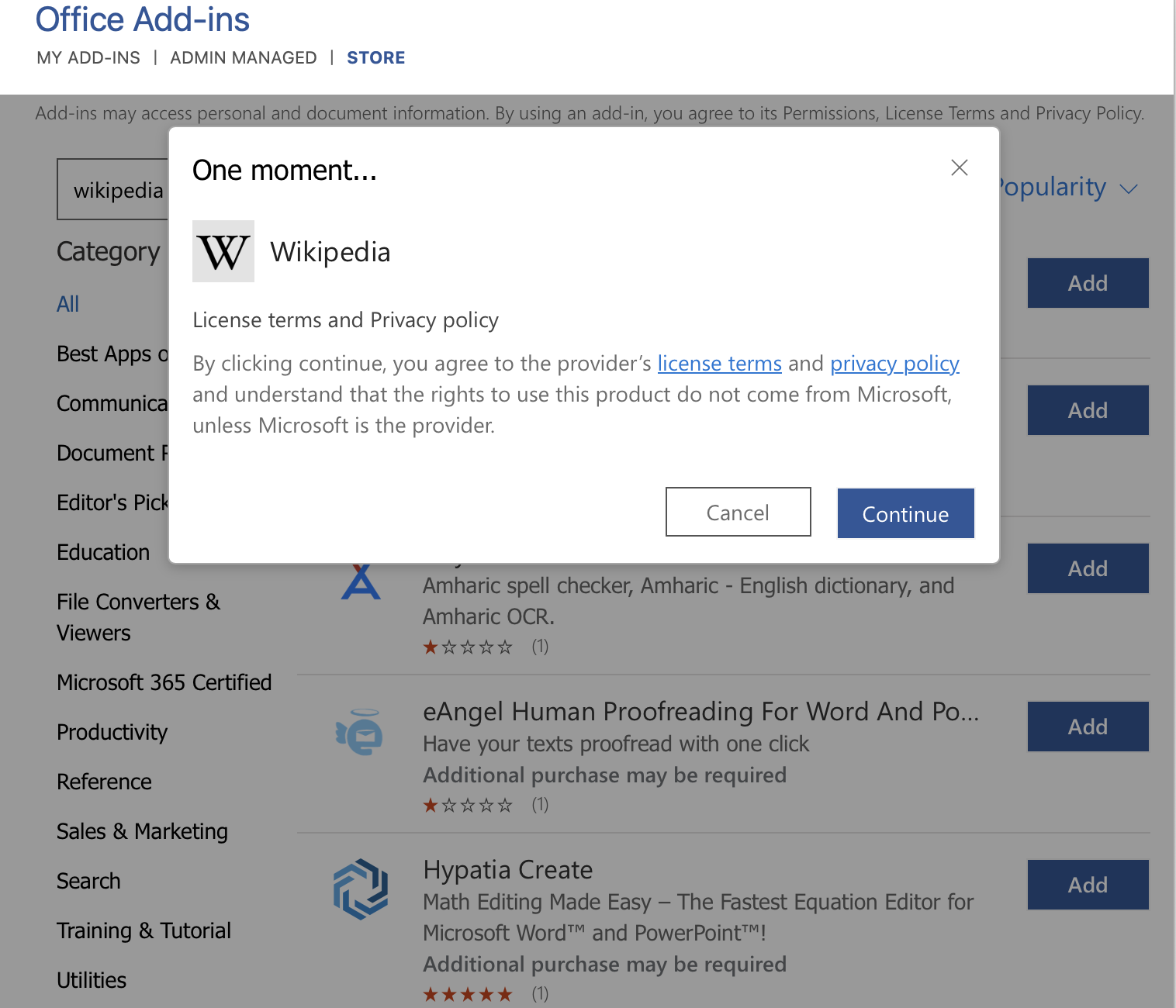}
    \caption{Installing the Wikipedia app on Microsoft Word.}
    \label{fig:wikpedia-word}
  \end{subfigure}
  \hfill
  \begin{subfigure}[b]{0.28\textwidth}
    \centering
    \includegraphics[width=\textwidth]{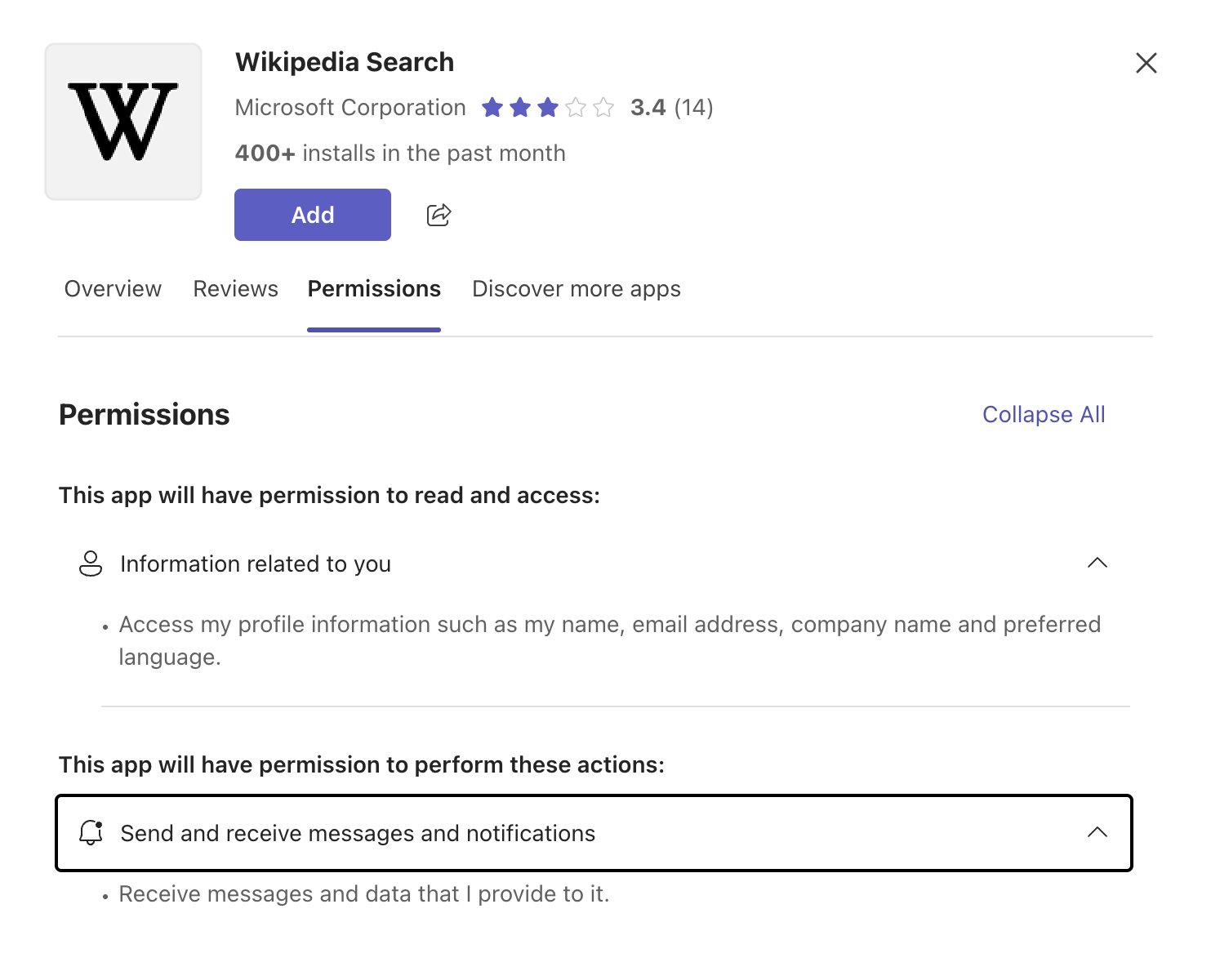}
    \caption{Installing the Wikipedia app on Microsoft Teams.}
    \label{fig:wikipedia-teams}
  \end{subfigure}
  \hfill
  \begin{subfigure}[b]{0.28\textwidth}
    \centering
    \includegraphics[width=\textwidth]{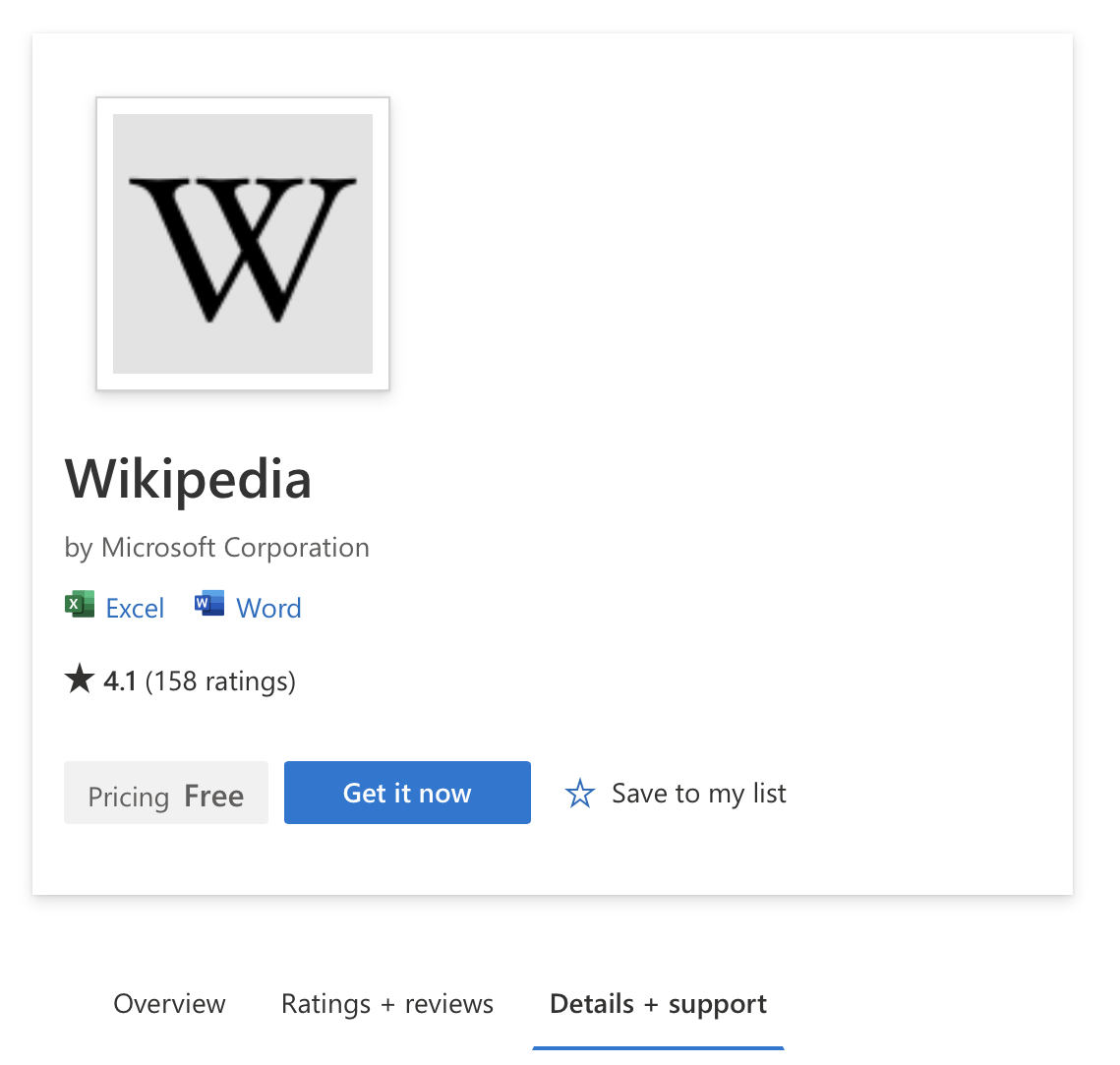}
    \caption{Installing the Wikipedia app on the Microsoft Marketplace.}
    \label{fig:wikipedia-market}
  \end{subfigure}
  
  \caption{Example installation interfaces for the same application across different M365 distribution channels, illustrating inconsistencies in permission disclosure.}
  \label{fig:wikipedia-installs}
\end{figure*}

In this section, we provide an overview of the quite complex Microsoft 365 ecosystem, what it offers and how it is structured. We introduce the concept of tenant, the applications within it and the permission model that governs the access to resources.

\subsection{Microsoft Entra ID and Tenants}

M365 services are organised into isolated environments called \emph{tenants}. A tenant represents a dedicated instance of Microsoft 365 associated with a specific organisation, containing its users, applications, and data. Each tenant is backed by a Microsoft Entra ID directory, which manages identities and authorisation.

When a developer registers an application globally, an \emph{application object} is created. Upon installation in a tenant, a corresponding \emph{service principal} is instantiated locally. Service principals therefore represent applications within a tenant. They act as the application's identity within that tenant and are the entities to which permissions are granted.

Permissions assigned to a service principal determine which of the tenant resources the application can access, including user-specific data and organisation-wide resources. As a result, analysing service principals and their associated permissions provides a direct view of the access surface introduced by third-party applications within a tenant. 

\subsection{OAuth Permission Model in M365}
\label{sec:microsoft_graph_permissions_m365}
Authorisation in M365 relies on the OAuth 2.0 framework, where applications obtain access tokens containing explicitly granted permissions (also referred to as \emph{scopes}). These permissions define the set of resources and operations an application is allowed to perform within a tenant.

M365 distinguishes three main types of permissions:
\textbf{Delegated permissions} allow an application to act on behalf of a signed-in user. In this case, the application's access is limited to the resources that the user is authorised to access.
\textbf{Application permissions} grant an application direct access to resources at the tenant level, independently of any signed-in user. These permissions typically require administrator consent and may enable organisation-wide access to sensitive resources such as directories, mailboxes, files, or chats.
\textbf{Resource-Specific Consent (RSC)} permissions provide scoped access to specific Teams resources (e.g., a particular team or chat), enabling finer-grained control compared to tenant-wide permissions. For instance, a person owning both team A and team B can consent to an application accessing only the resources of team A without affecting team B.

From a security perspective, application permissions are particularly sensitive, as they may enable persistent and tenant-wide access to organisational data. Analysing the type and scope of permissions requested by service principals, therefore, provides insight into the potential attack surface introduced by third-party applications.

 \subsection{Microsoft Graph and permission pattern}
\label{sec:permissions_pattern_m365}


Microsoft Graph is the primary API through which third-party applications access M365 resources.
These resources are locked behind authentication and role-based authorisation, which is managed through permissions. While OAuth standards do not define a strict pattern for naming permissions, the best practice is to use the pattern
<\texttt{resource}>.<\texttt{operation}>.<\texttt{constraint}>, where
\begin{itemize}
    \item \texttt{resource} refers to a Microsoft Graph resource to which the permission allows access (e.g., \texttt{User}, \texttt{Chat}, \texttt{Mail}, \texttt{Files}, ...);
    \item \texttt{operation} refers to the Microsoft Graph operations that are allowed on the data exposed by the resource (e.g., \texttt{Read}, \texttt{ReadWrite}, \texttt{ReadBasic}, \texttt{Create}, \texttt{Send}, ...);    
    \item \texttt{constraint} determines the potential extent of access an app has within the directory (e.g., \texttt{All}, \texttt{Directory}, \texttt{OwnedBy}, \texttt{Shared}, ...). This field is optional, and its default value is \texttt{OwnedBy}.
\end{itemize}

For example, \texttt{User.Read} allows an application to read the profile of the signed-in user, whereas \texttt{User.Read.All} enables reading the profile information of all users in the tenant. The \texttt{All} constraint therefore expands access from user-scoped to tenant-wide.

\subsection{Application Types and Distribution Channels}
Applications integrating with M365 can be broadly divided into two categories:
\begin{itemize}
    \item \textbf{Add-ins}: extend built-in services such as Teams, SharePoint, Outlook, or Word by adding UI elements, commands, tabs, or workspace-level functionality.
    \item \textbf{Third-Party Applications}: {externally-provided applications, often SaaS services, that} interact with M365 resources through Microsoft Graph APIs (e.g., project management platforms or calendar integrations).
\end{itemize}

{These categories are not mutually exclusive: an add-in may also request Microsoft Graph permissions and be represented in the tenant as a service principal.}

Applications can be installed through multiple distribution channels, including the official Teams and Office stores, and the global Microsoft Marketplace.\footnote{Microsoft calls \emph{Office add-ins} and \textit{Teams Apps} those app which extends Office or Teams products, respectively; the term \textit{Apps for Microsoft 365} is the generic term which includes all options.}
When installed, each application requests OAuth permissions according to its functionality.
One example is the ``Wikipedia'' app for Word and Excel,\footnote{\url{https://marketplace.microsoft.com/en-us/product/office/wa104099688}, accessed on \today.} that allows users to directly browse Wikipedia within Word or Excel.
In addition, a tenant or a user can distribute custom-developed app and offer them to users in other tenants. These custom app are not listed in any store.

At last, third-party apps can access M365 resources via API. For instance, Zoom or Notability can request to access Outlook Calendar or OneDrive files in the M365 cloud. No official store or repository exists, as any third-party app can use the Entra ID API, a subset of the Graph API. All apps are instantiated as service principals within an Entra ID tenant, as previously described.

Importantly, permission transparency and consent mechanisms vary across these channels. Some stores expose detailed permission lists before installation, while others provide limited or no visibility into the scopes that will be granted. This fragmentation complicates systematic auditing and motivates the need for an ecosystem-wide analysis of requested permissions.

It is important to note that the decision whether to allow or not the installation of an app in a tenant is totally delegated to the tenant administrator. Given the opaqueness and lack of information in the M365 app ecosystem, admins need tools to make informed decisions about authorising the use of an app.

%% file: text/3-dataset.tex
\section{Dataset collection}
\label{sec:dataset}

Given our goal of exploring the permissions apps ask for, we need at least three main features:
\begin{itemize}
    \item \textbf{App ID}: the app identifier and name.
    \item \textbf{Description}: a textual description of the app's purposes and functionalities, later used for topic modelling.
    \item \textbf{Set of Permissions}: the set of requested permissions, required for the AD stage.
\end{itemize}

\subsection{Challenges in dataset retrieval}

The M365 ecosystem does not provide a unified catalogue exposing all these attributes. Applications are distributed across multiple channels (e.g., Teams store, SharePoint store, Microsoft Marketplace, and direct third-party integrations), and each exposes different subsets of metadata. In particular, permission information is often incomplete, inconsistently disclosed, or accessible only after installation within a tenant.
Figure~\ref{fig:wikipedia-installs} shows three examples of installing the Wikipedia application in different contexts: Microsoft Word, Microsoft Teams, and the Microsoft Marketplace, which offer a version of the application compatible with Word and Excel. Each of the stores follows a different path for installing the app, and, most importantly, the permission-related information is not expressed (or present) in a consistent way, complicating a user's understanding of the privacy-related consequences of one's choice. For instance, among the three options, only Microsoft Teams (Figure~\ref{fig:wikipedia-teams}) provides the users with some information about the permissions that will be requested from the application.

\begin{figure}[t]
    \centering
    \includegraphics[width=.7\columnwidth]{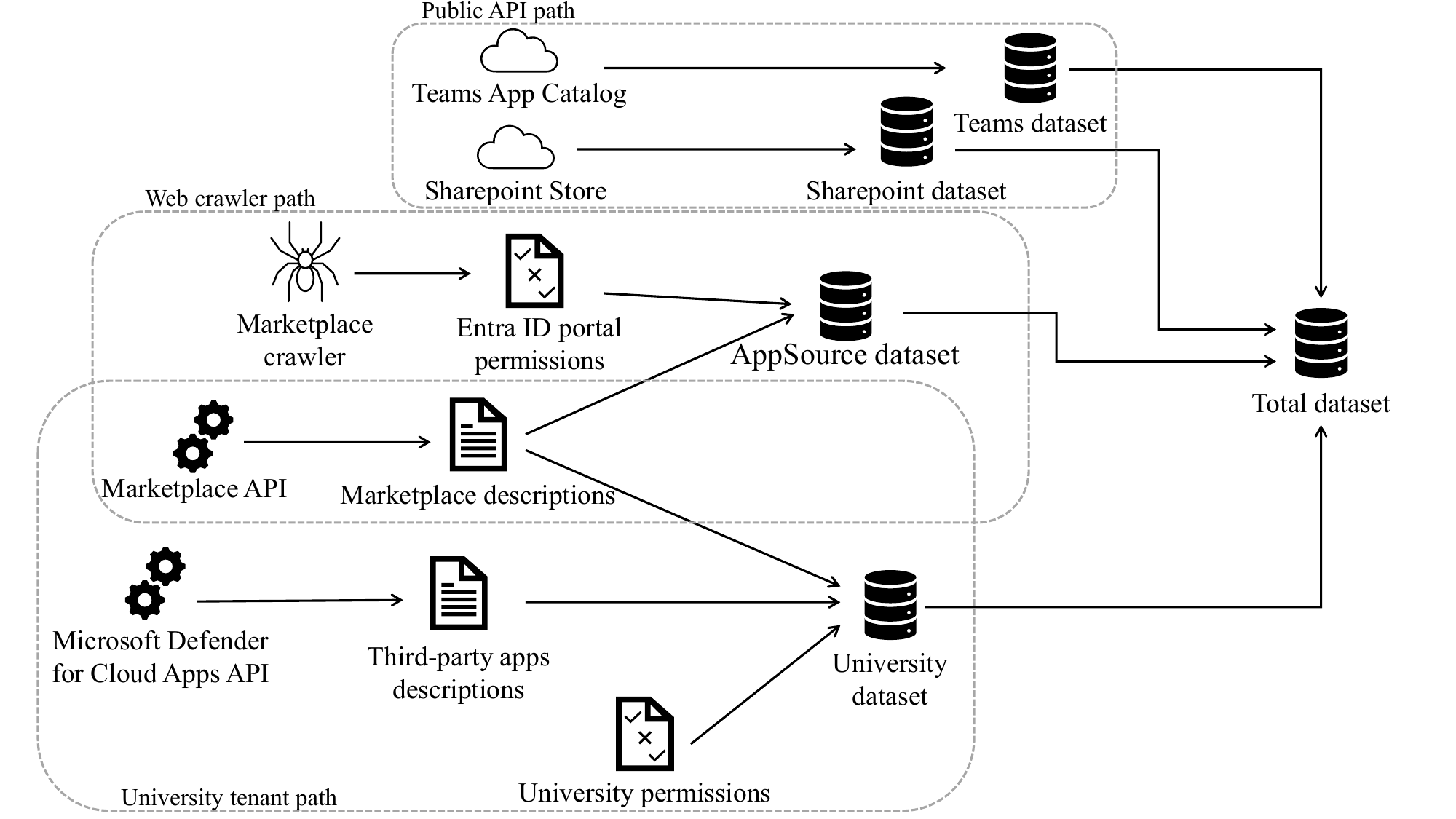}
    \caption{Overview of the data collection pipeline via public API extraction, tenant-side deployment, and operational tenant data integration.}
    \label{fig:dataset-paths}
\end{figure}

This fragmentation prevents direct large-scale measurement and necessitates a multi-source data collection strategy. We therefore combine public APIs, automated tenant-side deployment and scraping, and operational tenant data to construct the most comprehensive dataset currently attainable. Figure~\ref{fig:dataset-paths} summarises the different collection paths:
\begin{enumerate}
    \item We leverage public APIs that multiple app marketplaces (Marketplace, Teams Store, SharePoint Store, etc.) expose to retrieve\ the list of apps.

    \item We design a Web scraping tool that automatically accesses a test tenant, deploys apps, accepts installation prompts (if any) and extracts their properties through the Entra ID portal;

    \item We analyse apps that were installed in our University's operational tenant. The set includes some third-party apps that are not present in the other sources.
\end{enumerate}

Merging and consolidating the information from these different sources requires ingenuity, given that no unique app ID is available.

\subsection{Public APIs}
\label{subsec:public_apis}
We first leverage publicly available APIs exposed by official Microsoft application stores to retrieve application metadata at scale. In particular, we collect data from the Microsoft Teams app catalogue and the SharePoint store, which together host a substantial portion of M365 add-ins.

\textbf{Teams store}: The Microsoft Graph provides an officially documented API\footnote{\url{https://graph.microsoft.com/v1.0/appCatalogs/teamsApps?\$expand=appDefinitions}, accessed on \today} to access the Microsoft Teams app catalogue. Crawled in September 2025, it contains about 2,800 apps with descriptions, out of which only 347 contain a non-empty RCS permission field.

\textbf{SharePoint store}: When logged into SharePoint, a user can access the SharePoint Store\footnote{\url{https://<TENANT_NAME>.sharepoint.com/sites/appcatalog/_layouts/15/appStore.aspx/sharePointStore}, accessed on \today} to install apps. Given the app ID (obtained via the Marketplace API), it is possible to retrieve the list of requested permissions by calling an undocumented and authenticated API.\footnote{\url{https://<TENANT_NAME>.sharepoint.com/sites/appcatalog/_layouts/15/storefront.aspx?task=GetAppAdditionalDetails&bm=US&cm=en-US&appid=<APP_ID>&catalog=0&uirequest=1}, accessed on \today.} The API requires setting the right cookies within the request headers.
Following the crawl, 221 apps out of 1,222 have a non-empty permission field.

Across both sources, permission disclosure is partial and inconsistent: many applications provide descriptions but omit their requested OAuth scopes. This limitation prevents comprehensive ecosystem analysis using public APIs alone and motivates the need for additional collection strategies.

\subsection{Scraping Marketplace for Apps}

The Microsoft Marketplace exposes application descriptions but does not publicly disclose the full set of requested OAuth permissions. Since permission information becomes visible only after installation within a tenant, we implement an automated tenant-side deployment procedure to retrieve this data.

\subsubsection{Retrieving list of apps and descriptions}

First, we use undocumented but publicly accessible Marketplace endpoints to enumerate applications and retrieve their metadata (e.g., name and description). Crawling in September 2025 identifies 8,232 add-ins across M365 products (Table~\ref{tab:Marketplace_products}). The API returns the apps' description and additional information (e.g., rating, popularity scores, etc.). Yet, these endpoints do not expose permission information.


\begin{table}[H]
    \centering
    \caption{Marketplace applications per M365 product category. The total is lower than the sum of individual counts because some applications are listed under multiple products.}
    \begin{tabular}{lr}
        \toprule
        \textbf{Product} & \textbf{Total Apps} \\
        \midrule
        Excel      & 1,338 \\
        Office     & 412  \\
        Outlook    & 2,069 \\
        PowerPoint & 480  \\
        SharePoint & 1,222 \\
        Teams      & 2,595 \\
        Word       & 1,306 \\
        \midrule
        
        \textbf{Total (unique)}      & \textbf{8,232} \\
        \bottomrule
    \end{tabular}
    \label{tab:Marketplace_products}
\end{table}

\subsubsection{Collecting permissions from a tenant}

To obtain permissions, we create a dedicated test tenant and implement an automated deployment tool using Selenium WebDriver. For each application, the tool initiates installation via the Marketplace interface, completes authentication with administrative credentials, accepts consent prompts when present, and extracts the granted permissions from the Entra ID admin portal. This process allows us to retrieve the effective OAuth scopes associated with each service principal.\footnote{The crawling tool's source code will be available upon acceptance.}

The tool handles most of the exceptions that could happen during the deployment process, such as the absence of the OAuth prompt, timeouts, generic errors, etc. Despite this, the crawler successfully deployed 432 apps listed in Marketplace (about 4\% of the total). This low success rate is primarily due to (i) the OAuth consent unexpectedly failing during deployment (further attempts do not help)\footnote{We frequently encountered this error also during manual deployment.}; (ii) apps not asking for permissions when deployed (useless for our purpose); (iii) the deployment process of the app requiring additional steps (e.g., registration through the app portal).

We acknowledge that, for some applications, automatically installing them may be against their terms of service. However, we only install the applications without ever using them, restricting the impact of the automated data extraction only to Microsoft servers, for which the few hundred of downloaded applications represent a negligible load.

\subsubsection{Linking descriptions and permissions}

Applications extracted from the Marketplace APIs and those observed in the Entra ID portal may differ in naming conventions. For example, an application listed in the Marketplace under a generic name may appear within a tenant with product-specific suffixes (e.g., “for M365” or “– Teams”), resulting in distinct service principal names.

To associate application descriptions with their corresponding permission sets, we adopt a two-stage matching strategy. First, we attempt an exact match between the app names and service principal display names. When no exact match is found, we apply a controlled partial matching procedure based on manually curated regular expressions that normalise common suffixes and prefixes.

To minimise incorrect associations, we inspect all ambiguous matches and discard cases where multiple candidate associations cannot be resolved confidently. Applications with uncertain mappings are excluded from the dataset. After merging crawler-derived permissions with metadata from other sources, the crawler contributes 330 unique applications. While deployment success is limited to a subset of Marketplace applications, successfully deployed apps span multiple product categories and exhibit heterogeneous permission counts, with coverage across diverse app types.

This conservative approach prioritises precision over recall in linking descriptions to permissions, reducing the risk of misattributing permission sets to incorrect applications.


\subsection{University tenant apps}

To complement the Marketplace-based collection, we analyse applications installed in our University's operational M365 tenant, comprising more than 35,000 users who installed applications in the tenant. The file represents the snapshot in May 2025.\footnote{In mid-2025, Microsoft updated its M365 security baseline, switching the third-party application consent policy from user-level opt-out to admin-required opt-in to prevent unauthorized data access. Since then, users cannot install any new application.} This source captures both add-ins and third-party applications actively deployed in a real-world enterprise environment, including integrations not listed in official stores.

University IT administrators provided an anonymised export of service principals present in the tenant. The dataset includes 285 applications along with their associated permission sets. Because the Entra ID interface does not expose application descriptions, we retrieve corresponding descriptions by querying Marketplace metadata and, when necessary, the Microsoft Defender for Cloud Apps API.

As before, naming discrepancies between service principals and Marketplace listings require controlled matching. We apply the same two-stage matching strategy and manually validate ambiguous cases to ensure correct associations between descriptions and permission sets. After filtering unmatched or ambiguous entries, this source contributes 285 applications with both descriptions and permissions.\footnote{Permissions observed in an operational tenant may reflect administrator-imposed restrictions and therefore represent an effective permission configuration rather than the full set declared by the developer. In our case, all apps were allowed with their default permissions.}



\subsection{The final dataset}
We merge applications collected from public APIs, Marketplace deployment, and the operational tenant into a unified dataset. Because no global identifier exists across distribution channels, consolidation requires careful deduplication and validation.

\begin{table}
    \centering
    \caption{Number of applications collected from each source prior to deduplication.}
    \begin{tabular}{lr}
        \toprule
        \textbf{Source} & \textbf{\# of Apps} \\
        \midrule
        Crawler    & 432 \\
        University & 285 \\
        SharePoint & 221 \\
        Teams      & 348 \\
        \midrule
        \textbf{Total}      & \textbf{1,286} \\
        \bottomrule
    \end{tabular}
    \label{tab:apps_per_source}
\end{table}

\paragraph{Handling duplicates}
Given that no global identifier exists, we check for possible duplicates using the app's official name and that of the service principal:
for apps appearing in the Teams and the crawler data (49 apps), we merge permission information as the RCS and tenant-wide permissions are complementary. 
For duplicates in the Teams and the University data (19 apps) and in the crawler and University data (28 apps), we keep the Teams and crawler permissions that are more complete (University data may reflect tenant-level filtering).

\paragraph{Handling incorrect matches}
We next look for discrepancies that could have been introduced during the ``direct-then-partial'' match approach 
used to complete the description field. Specifically, we look for those apps with the same title (i.e., the same manifest) but different appName (i.e., different Service Principals). These cases may happen because the same apps might trigger the creation of multiple Service Principals (e.g., Zoom $\rightarrow$ ``Zoom for M365'' and ``Zoom for Teams''). After manually checking all these candidates, we drop 61 apps with ambiguous associations.

\paragraph{Description cleaning}
Finally, we perform a basic pre-processing step consisting of filtering out apps with very short descriptions (i.e., less than 10 words) or with non-English descriptions. We also clean the description text by removing HTML tags and non-alphanumeric characters. Table~\ref{tab:preprocessing_summary}  summarises the pre-processing steps and the final number of apps we use for our analysis.

\begin{table}
    \centering
    \caption[Preprocessing Summary]{Summary of dataset consolidation and preprocessing steps, showing the resulting application count at each stage.}
    \begin{tabular}{lr}
        \toprule
        \textbf{Preprocessing Step} & \textbf{\# of Apps} \\
        \midrule
        Initial number of apps with descriptions & 1,286 \\
        Handling duplicates & $-$96 \\
        Handling discrepancies & $-$61 \\
        Filtered out (not in English) & $-$40 \\
        Filtered out (less than ten words) & $-$20 \\
        \midrule
        \textbf{Final number of apps} & \textbf{1,069} \\
        \bottomrule
    \end{tabular}
    \label{tab:preprocessing_summary}
\end{table}

Overall, the coexistence of multiple app stores, incomplete metadata, the absence of a unified management interface, and non-standardised Service Principal names permits us to retrieve information for only 1,069 out of more than 8,000 applications, highlighting the current fragmentation and limited maturity of the M365 app ecosystem---see Table~\ref{tab:preprocessing_summary}.

\subsection{Dataset characterisation}

\begin{figure}
    \centering
    \includegraphics[width=.6\columnwidth]{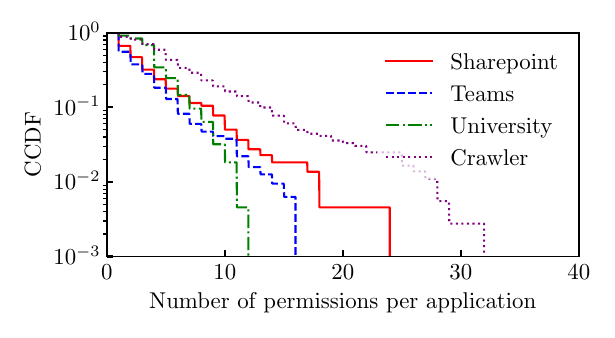}
    \caption{CCDF of the number of requested permissions per application, stratified by data source.}
    \label{fig:ccdf-per-source}
\end{figure}

We analyse the distribution of requested permissions across the 1,069 applications in the final dataset. Figure~\ref{fig:ccdf-per-source} reports the complementary cumulative distribution function (CCDF) of the number of permissions requested per application, stratified by data source.
The distribution exhibits substantial variability both across and within sources. Applications obtained via Marketplace deployment display a heavier tail, with several apps requesting more than 20 permissions. In contrast, applications observed in the operational tenant tend to request fewer permissions, likely reflecting administrator-level filtering or differences in app type. Across all sources, permission counts range from 1 to 32 per application.

\begin{figure}
    \centering
    \includegraphics[width=.6\columnwidth]{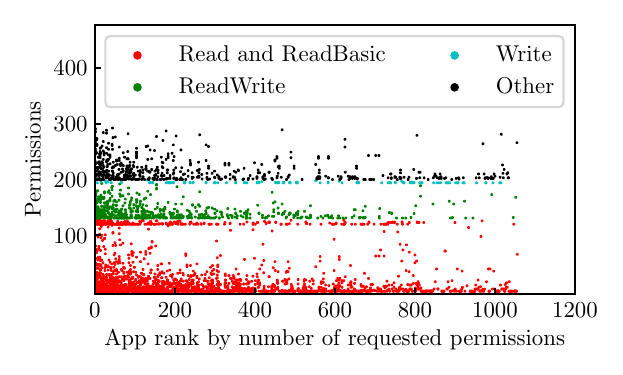}
    \caption{Permission–application matrix. Applications (x-axis) are ordered by decreasing number of requested permissions; permissions (y-axis) are grouped by operation type.}
    \label{fig:scatter-app-permission}
\end{figure}

In total, we observe more than 300 distinct OAuth permissions. Some permissions are highly prevalent: for example, 48\% of applications request \texttt{User.Read}, 37\% request \texttt{openid}, and 36\% request \texttt{profile}. Conversely, a large number of permissions appear only in a small fraction of applications.

Figure~\ref{fig:scatter-app-permission} visualises the full permission–application matrix. Applications are ordered by decreasing number of requested permissions, while permissions are grouped by operation type (e.g., \texttt{Read}, \texttt{ReadWrite}, ...). The figure highlights both highly permission-intensive applications and others requesting rare or uncommon scopes.
Rarer permissions are requested by the more permission-demanding apps (dots in the leftmost part of the plot). Interestingly, some apps request very few but rarely requested permissions (dots in the rightmost part of the plot). 

This heterogeneity suggests that applications differ significantly in their effective privilege profiles. In particular, some applications request extensive combinations of permissions, while others request only a minimal subset. We hypothesise that part of this variability reflects differences in declared functionality, motivating a topic-aware analysis of permission patterns.

%% file: text/4-clustering.tex
\section{Topic retrieval}
\label{sec:clustering}

Given the heterogeneity of apps and permissions, we proceed by clustering apps based on their purpose. For this task, we use the applications' description and rely on state-of-the-art topic modelling to group apps into clusters. We use BERTopic \cite{BERTopic_Grootendorst_2022}, one of the most recent, scalable and effective models for topic extraction, leveraging transformer-based embeddings.

\subsection{Topic retrieval methodology}
\label{sec:topic-retrieval}

BERTopic is organised in a six-block pipeline:
\begin{enumerate}
    \item \textbf{Embedding Model}: the first step consists of the extraction of document embeddings. The default embedding model is Sentence-BERT~\cite{reimers2019sentence}, which is highly capable of capturing the semantic similarity among documents. This step maps each document in the embedding space where semantically similar (different) documents are projected nearby (far apart).
    \item \textbf{Dimensionality Reduction}: As the embedding space is usually high-dimensional, BERTopic reduces the representation space dimensionality to avoid the curse of dimensionality problem during the clustering step. BERTopic uses UMAP~\cite{sainburg2021parametric}, one of the most effective algorithms in preserving both the local and global high-dimensional space in lower dimensions.
    \item \textbf{Clustering Algorithm}: To find clusters of documents with similar topics, BERTopic uses HDBSCAN~\cite{mcinnes2017hdbscan}, a parameter-free and effective algorithm to capture clusters of varying densities. The results are clusters of documents that are semantically similar.
    \item \textbf{Word Vectorisation}: To characterise each topic, BERTopic adopts a bag-of-words representation. All documents within a cluster are combined, and each word is represented as a feature whose value corresponds to its frequency within that cluster. We employ the \texttt{CountVectorizer} tool to compute word occurrence counts, which are later used to identify the most representative terms for each topic.
    \item \textbf{Topic Representation}: The final step is to extract the most characterising words for a topic, i.e., words that are frequent in such a topic, and rare in others. BERTopic uses a class-based Term Frequency-Inverse Document Frequency (c-TF-IDF) for this. The result is the extraction of keywords that well represent each topic.
    \item \textbf{Topic Fine-Tuning}: Optionally, one can further refine the topic representation by using other Natural Language Processing techniques. We do not proceed with this step.
\end{enumerate}

All the blocks composing the BERTopic pipeline require tuning of their own hyperparameters. The objective is to extract the most coherent topics possible from the descriptions of the applications in the dataset. To achieve this, we follow a structured cascade approach for hyperparameter tuning: for each model in the pipeline (i.e., SBERT, UMAP, HDBSCAN, CountVectorizer, c-TF-IDF), we select the best hyperparameters and the possible values to test. For each block, we select the best-performing configuration based on objective metrics such as the clusters' silhouette scores and semantic coherence. Having fixed the best parameters, we move on to the next block. We repeat each experiment using three different seeds to ensure stability of the results and robustness against randomness. We detail the BERTopic final parameters and the search space values in Appendix~\ref{sec:bertopic-parameter-selection}.

\subsection{Retrieved topics}

We run BERTopic using the description of each app as input documents. As output, BERTopic returns clusters of apps whose descriptions are semantically similar, along with the top-5 keywords that better characterise their topic extracted using the c-TF-IDF scores. We identify 24 clusters, which we manually name in a more descriptive form based on the returned keywords. {As the keywords only serve to provide an intuition for the cluster app goals, we consider 5 keywords as a plausible trade-off between synthesis and semantic richness.} 

The topics reflect the nature of M365 apps and services, with the most popular ones related to productivity and collaboration. Clusters contain a variable number of apps, the largest one containing 120 of them ("File management" topic), 40\% of them containing more than 50 applications, 30\% containing less than 30 apps.
Manual inspection indicates coherent separation of applications into semantically meaningful topics.
Here, we provide a qualitative interpretation of the top four and the bottom four clusters by number of applications.  For completeness, we report the complete list, their top-5 keywords, description, and number of applications in Appendix~\ref{sec:extracted-topics}.

The four most popular topics revolve around the following functions:
\begin{itemize}
    \item \textbf{File Management, 122 apps}: applications related to file storage, sharing, and handling;
    \item \textbf{Security, 75 apps}: applications focused on asset management and protection;
    \item \textbf{Notifications, 65 apps}: applications that provide notifications and alerts to users;
    \item \textbf{E-learning, 65 apps}: applications for online learning and educational purposes.
\end{itemize}

On the other hand, the least popular topics are more specific and represent niche or innovative areas of the M365 application world in our dataset. These are:
\begin{itemize}
    \item \textbf{Human Resources, 20 apps}: applications focused on human resources management, recruitment and life cycle handling;
    \item \textbf{AI Translation, 19 apps}: applications focused on translation services using artificial intelligence;
    \item \textbf{PDF Management, 15 apps}: applications focused on PDF document management and manipulation;
    \item \textbf{AI Legal Assistant, 12 apps}: applications designed for legal assistance and document review using artificial intelligence.
\end{itemize}

\subsection{Topics and permissions}
\label{subsubsec:results_topics_similarity}

To analyse how permissions vary across functional categories, we construct a permission profile for each topic. Let $\mathcal{P}$ denote the set of all distinct permissions observed in the dataset. For each topic $i$, we define a vector $P_i \in \mathbb{R}^{|\mathcal{P}|}$, where each component corresponds to a permission $p \in \mathcal{P}$.

The term frequency (TF) of permission $p$ in topic $i$ is defined as the proportion of applications in topic $i$ requesting $p$. To emphasise permissions that are distinctive to specific topics, we apply an inverse document frequency (IDF) weighting computed across topics. The resulting TF-IDF representation mitigates the effect of topic size and down-weights globally common permissions.
We compute pairwise cosine similarity between topic vectors:
\[
s(i,j) = \frac{P_i \cdot P_j}{\|P_i\|\|P_j\|},
\]
where higher values indicate more similar permission profiles.

\begin{figure}
    \centering
    \includegraphics[width=.6\columnwidth]{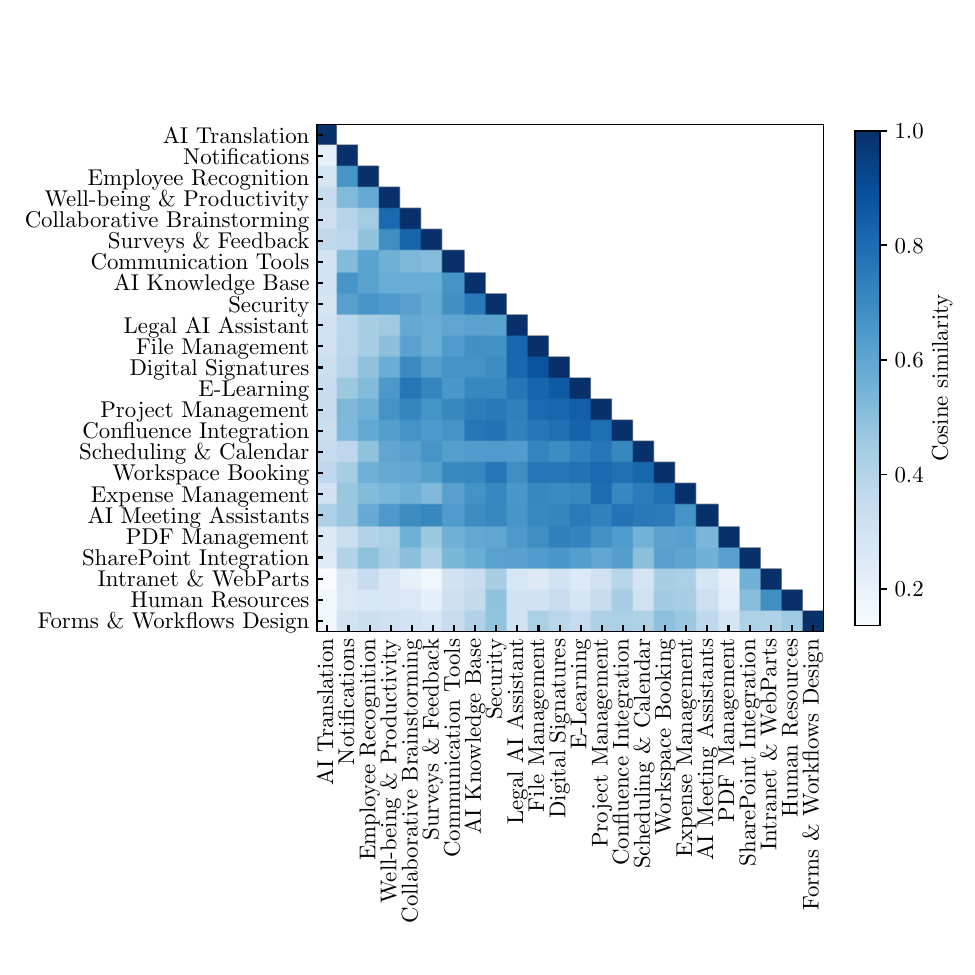}
    \caption[Heatmap of Topics Permissions Similarities]{Pairwise cosine similarity between topic-level permission profiles computed from TF-IDF weighted permission frequency vectors.}
    \label{fig:topics_permissions_similarities}
\end{figure}
Figure~\ref{fig:topics_permissions_similarities} reports the resulting similarity matrix. While some topics exhibit overlapping permission patterns—reflecting shared resource requirements—others display markedly distinct profiles (e.g., “Intranet \& WebParts” and “AI Translation”), suggesting functional differentiation in access needs.

Overall, apps with different purposes request distinct permission sets, confirming that permission profiles mirror functional goals.



\begin{figure}
    \centering
    \begin{subfigure}{0.45\columnwidth}
        \centering
        \includegraphics[width=\columnwidth]{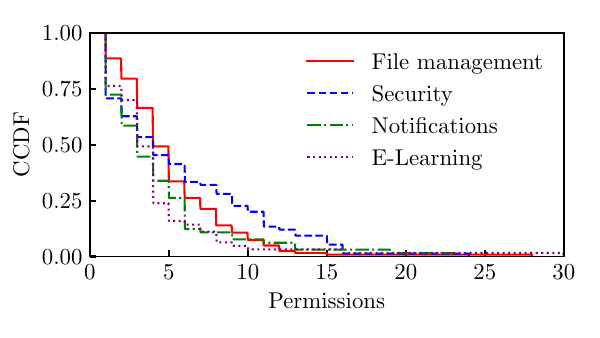}
        \caption{Five larger topics.}
        \label{fig:ccdf-per-topic-top}
    \end{subfigure}
    \hspace{.05\columnwidth}
    \begin{subfigure}{0.45\columnwidth}
        \centering
        \includegraphics[width=\columnwidth]{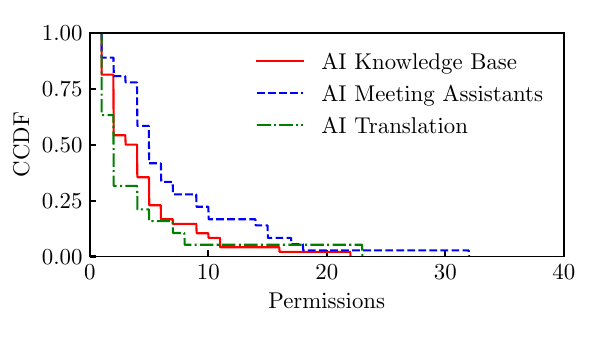}
        \caption{Three topics related to Artificial Intelligence.}
        \label{fig:ccdf-per-topic-ai}
    \end{subfigure}

    \caption{Complementary CDFs for the number of permissions requested by applications within selected topics.}
    
    \label{fig:ccdfs-per-topic}
\end{figure}

To further examine intra-topic variability, Figure~\ref{fig:ccdfs-per-topic} reports the complementary cumulative distribution functions (CCDFs) of the number of permissions requested by applications within selected topics. Even among applications serving similar functional purposes, we observe considerable dispersion: some apps request only a small number of permissions, while others request more than 20. This pattern holds across both general productivity topics and AI-related clusters.

Figure~\ref{fig:ccdfs-per-topic} shows the distribution of permissions within the same topic, and across different---but related---topics. In Figure~\ref{fig:ccdf-per-topic-top}, we compare the permission distribution for the four most popular topics. Security-related apps request a larger share of permissions. All the topics present at least one application that requests more than 20 permissions. The same pattern emerges by comparing applications offering AI-driven services---see Figure~\ref{fig:ccdf-per-topic-ai}. ``AI Meeting Assistants'' tend to request more permissions, with $\approx$20\% of them requesting more than 10 permissions. Curiously, almost 40\% of ``AI Translation'' apps require just one permission, mostly \texttt{OnlineMeetingParticipant.ToggleIncomingAudio.Chat}. Yet other apps that offer similar functionalities request 10-20 permissions.

These results show that applications requesting a high number of permissions appear in every topic, while other applications, on the other hand, require very few permissions. This high disparity suggests the possible presence of outliers inside each topic.

\begin{figure}
    \centering
    \includegraphics[width=.7\linewidth]{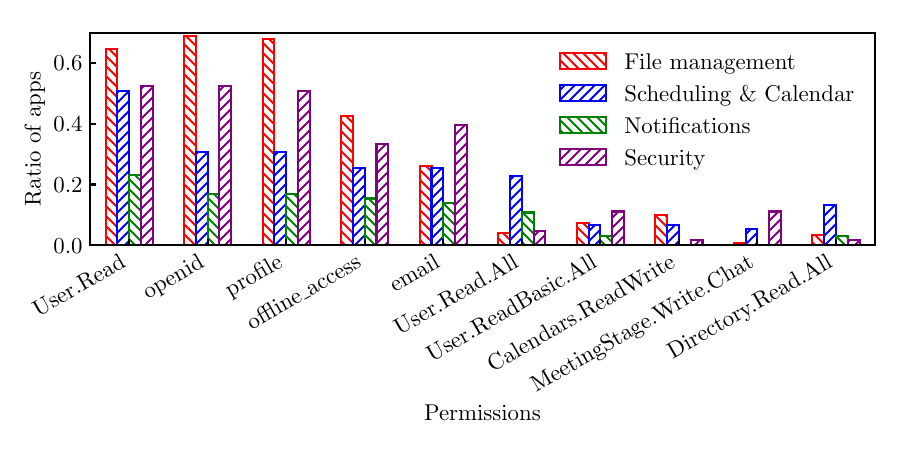}
    \caption{Fraction of applications within selected topics requesting each of the ten most globally frequent permissions.}
    \label{fig:top-permissions}
\end{figure}

To further corroborate the assumption that different topics require different permission sets, we plot in Figure~\ref{fig:top-permissions} the percentage of applications inside each topic that request a given permission. We select the four topics with the largest number of applications. We consider the top-10 permissions according to their global popularity inside our dataset. For each topic, we report the ratio of apps requesting such a permission. Different patterns exist for different topics. For instance, File management applications tend to request more permissions about \texttt{User.Read}, \texttt{openid}, and \texttt{profile} with respect to the other topics. On the other hand, Notification applications require a smaller share of the top permissions. At last, we observe that only a small percentage of apps request less popular topics (e.g., \texttt{Calendars.ReadWrite} and \texttt{MeetingStage.Write.Chat}).

Taken together, these observations suggest that permission usage exhibits both inter-topic structure and intra-topic variability. While functional similarity explains part of the permission patterns, the presence of significant deviations within topics motivates a topic-aware anomaly detection approach to identify applications whose permission profiles diverge from their functional baseline.

%% file: text/5-anomaly-detection.tex
\section{Anomaly detection}
\label{sec:anomaly-detection}

Our final objective is to identify applications whose requested permissions deviate from those of other applications serving similar functional purposes. 

\subsection{Anomaly detection methodology}
\label{sec:ad-method}

We employ three complementary unsupervised anomaly detection algorithms: One-Class SVM (OC-SVM), Local Outlier Factor (LOF), and Isolation Forest (iForest)~\cite{ChandolaAD2009}. OC-SVM models a global decision boundary enclosing the majority of samples; LOF detects local density deviations relative to neighbouring samples; iForest isolates anomalous points through recursive random partitioning. Using multiple paradigms reduces reliance on a single notion of abnormality and improves robustness.
Each algorithm assigns an anomaly score to each sample, allowing us to rank them from the most to the least anomalous.

Each application is represented as a binary vector over the permission vocabulary $\mathcal{P}$, where each dimension indicates the presence or absence of a given OAuth scope. This representation captures the effective access profile granted to the application's service principal. We perform anomaly detection separately within each semantic topic, under the assumption that applications with similar declared functionality should exhibit comparable permission patterns. Detecting anomalies across the entire dataset without topic separation would confound functional heterogeneity with abnormal behaviour.

Because no ground truth labels exist for anomalous applications, we adopt a controlled synthetic outlier injection strategy to tune hyperparameters. For each topic, we randomly select 20\% of applications and perturb their permission vectors by (i) removing $N$ common permissions among the top 30 most frequent in the topic and (ii) adding $N$ rare permissions among the 30 least frequent, with $N=1$ (hard), $2$ (medium), $4$ (easy).

\begin{figure}
    \centering
        \centering
        \includegraphics[width=0.6\columnwidth]{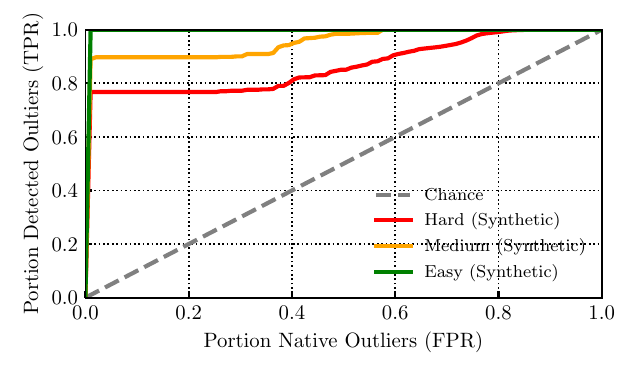}
        \caption{Average ROC curves for OC-SVM with synthetic outliers injection across all topics.}
        \label{fig:ocsvm_roc_curve_injected_outliers_synthetic}
\end{figure}

This procedure simulates deviations from typical permission usage while preserving realistic permission structures. We evaluate model performance using ROC curves and Area Under the Curve (AUC), treating perturbed applications as positives and unmodified ones as negatives. Experiments are repeated across $k=5$ folds to mitigate randomness. For each model, we select hyperparameters that maximise average AUC across topics under the medium-difficulty setting.
Figure~\ref{fig:ocsvm_roc_curve_injected_outliers_synthetic} shows the results for the OC-SVM algorithm, which is particularly good in identifying those apps for which we altered their permissions (being those global outliers in fact). For the sake of space, we detail all the results of the tuning phase in Appendix~\ref{sec:anomaly-detection-model-tuning}. This tuning procedure provides a principled way to calibrate detection models in the absence of labelled anomalies while maintaining comparability across topics.

\subsection{Anomaly detection results}


We apply each algorithm to identify the outliers in each topic.
Each AD algorithm produces a set of applications identified as anomalous. To increase robustness and reduce model-specific biases, we define the final set of anomalous applications as the intersection of the three sets.
This consensus-based approach ensures that only applications consistently flagged as anomalous across all detection methods are retained for further analysis. 




\begin{table}
  \centering
  \caption{Pairwise agreement between anomaly detection models, measured as intersection-over-union of flagged apps.}
  \label{tab:pairwise_agreement}
  \begin{tabular}{llc}
    \toprule
     &  & \textbf{Agreement} \\
    \midrule
    OC\text{-}SVM & LOF     & 0.77 \\
    OC\text{-}SVM & iForest & 0.74 \\
    LOF           & iForest & 0.84 \\
    \hline
    \multicolumn{2}{c}{\textbf{Overall}} & \textbf{0.67} \\
    \bottomrule
  \end{tabular}
\end{table}

Table~\ref{tab:pairwise_agreement} shows the pairwise agreement rates, computed as the proportion of applications classified identically by both models over the total number of anomalies flagged (intersection over union). Overall, there is substantial overlap among the models’ predictions, ranging from a 74\% agreement between OC-SVM and iForest to 84\% between LOF and iForest.
At last, we verify the global agreement score as the arithmetic mean of the pairwise Cohen’s kappa coefficients, following the approach proposed in~\cite{hallgren_2012_computing} for scenarios involving three raters evaluating the same set of items. The overall Cohen’s kappa coefficient across the three models is 0.48 on a scale from –1 to 1. Values around 0.5 are generally regarded as indicative of a good level of agreement among raters. The relatively high pairwise agreement rates and the strong global agreement score suggest that the models capture consistent notions of deviation.


Across all topics, 139 applications (13\% of the dataset) are flagged as anomalous under the consensus criterion. When anomaly detection is performed without topic separation, substantially fewer applications are identified, primarily those exhibiting extreme permission counts. This contrast highlights the importance of modelling permission profiles relative to functional peers rather than across heterogeneous application categories.

We emphasise that anomalous classification does not imply malicious intent. Rather, flagged applications exhibit permission profiles that significantly deviate from those of other applications within the same semantic topic. Such deviations may correspond to over-privileged configurations, under-privileged configurations, or legitimately distinct design choices. To better understand the nature of these deviations, we perform qualitative validation through LLM-assisted analysis and blind manual inspection.

\subsection{Finding anomalous permissions}
\label{sec:anom-permission}

To better understand why an application is identified as anomalous, we perform a topic-aware analysis of permission anomalies. Since the set of 139 anomalous applications was obtained by combining the outputs of multiple anomaly detection algorithms, merging their individual explanation mechanisms would have produced results that are difficult to interpret. Instead, we introduce a simple and interpretable \emph{surprise score} for each permission, quantifying how unusual the application's permission request is with respect to the typical behaviour of applications belonging to the same topic. For each permission~$j$, we compute the score:

$$S=\frac{x_j \log p_j + (1-x_j) \log (1-p_j)}{-\log(N)},$$

where $x_j$ indicates whether the target application requests permission~$j$, $p_j$ is the probability that permission~$j$ is requested by applications belonging to the same topic as the target application (i.e., the fraction of applications in that topic requesting the permission), and $N$ is the number of applications in the topic.

The most anomalous case occurs when all applications in a topic request a permission while the target application does not; or vice versa. In this situation, the numerator equals $-\log N$. Dividing by this quantity normalizes $S_j$ to the range $[0,1]$, independently of the number of applications in the topic, where a value of 1 denotes the highest degree of anomaly.

%% file: text/6-validation.tex
\newcommand{\IncoherentRisky}{\textit{Incoherent, Risky}\xspace}
\newcommand{\CoherentRisky}{\textit{Coherent, Risky}\xspace}
\newcommand{\CoherentSafe}{\textit{Coherent, Safe}\xspace}
\newcommand{\Insufficient}{\textit{Insufficient}\xspace}

\section{Qualitative verification}
\label{sec:validaton}


Our anomaly detection methodology leaves one important question open: are anomalous applications associated with higher security and privacy risks? To address this question, we use an LLM to provide a scalable, independent assessment of the alignment between application descriptions and requested permissions. Specifically, we analyse the most and least anomalous applications in each topic-specific anomaly ranking. We then manually inspect a subset of the most anomalous applications to evaluate the potential risks associated with their requested permissions.

Our assessment is purely static and is based on the inherent risk of the permissions requested by each application. Dynamic analysis of how permissions are actually used at runtime is beyond the scope of this work and is left for future research. 
\subsection{LLM validation}

To obtain a scalable semantic assessment of permission–description alignment, we employ GPT-5.2 via API as an auxiliary evaluation tool. For each application, we provide the LLM with its Marketplace description and full list of requested OAuth permissions, and prompt the LLM to classify the alignment between declared functionality and requested scopes.

\begin{framed}
\textit{Choose '\textbf{Coherent, Safe}' if the permissions are coherent with the description and do not raise any security concerns.}

\textit{Choose '\textbf{Coherent, Risky}' if the permissions are somewhat coherent with the description but raise some security concerns.}

\textit{Choose '\textbf{Incoherent, Risky}' if the permissions are not coherent with the description and raise significant security concerns.}

\textit{Choose '\textbf{Insufficient}' if the permissions are not sufficient to accomplish the intents declared in the description.}
\end{framed}

We evaluate applications at the tails of the anomaly distribution. For each anomaly detection model and each topic, we select the top 5\% most anomalous and top 5\% least anomalous applications based on per-topic anomaly scores. We also compute an aggregated ranking by summing the ranks across models and select the corresponding extremes. This strategy ensures coverage across topics and across different notions of deviation.

\begin{figure}
    \centering

    \begin{subfigure}{.45\columnwidth}
        \centering
        \includegraphics[width=\columnwidth]{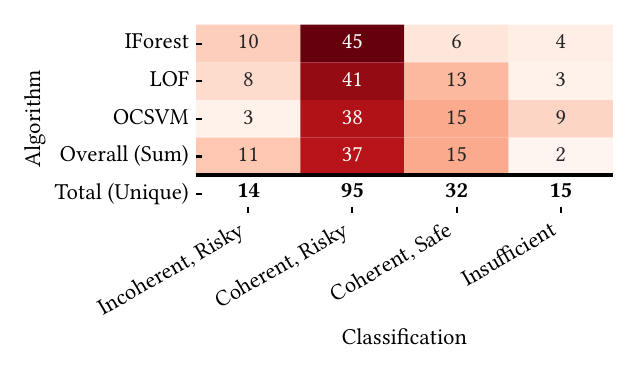}
        \caption{Breakdown of the most anomalous apps.}
        \label{fig:llm-result-heatmap-anomalous}
    \end{subfigure}
    \hspace{.05\columnwidth}
    \begin{subfigure}{.45\columnwidth}
        \centering
        \includegraphics[width=\columnwidth]{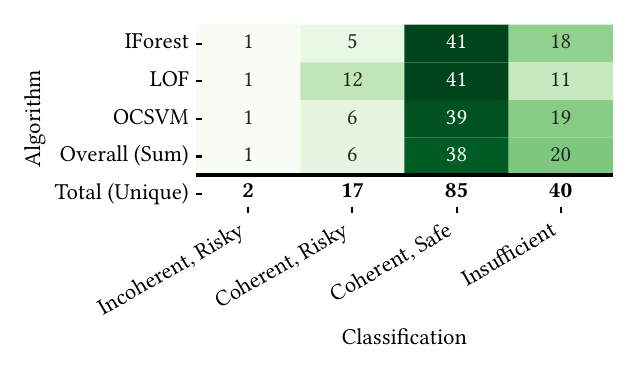}
        \caption{Breakdown of the least anomalous apps.}
        \label{fig:llm-result-heatmap-safe}
    \end{subfigure}

    \caption{Breakdown of LLM classification outcomes, for each anomaly detection model and for the aggregated ranking.}
    \label{fig:llm-result-heatmaps}
\end{figure}

For each anomaly detection algorithm rank, we select the 65 most-anomalous and 65 least-anomalous applications (after rounding per-topic quotas). Given description and permissions, the LLM returns both a categorical label and a short textual justification.
Figure~\ref{fig:llm-result-heatmaps} summarises the distribution of LLM classifications. The last row of the heatmaps details the number of unique applications with the given classification across all three algorithms and the rank sum.
Applications ranked among the least anomalous are predominantly labelled \CoherentSafe, whereas those ranked among the most anomalous are primarily labelled \CoherentRisky or \IncoherentRisky. This pattern is consistent across detection models and the aggregated ranking.

Notably, many anomalous applications are classified as \CoherentRisky. Inspection of the LLM explanations indicates that these cases typically involve permissions that are functionally plausible but excessively broad (e.g., tenant-wide \texttt{*.Read.All} or \texttt{*.ReadWrite.All} scopes).
For instance, \textit{Mindomo 9.0}, a collaborative mind mapping and brainstorming application, contains the \texttt{Files.ReadWrite.All} permissions. While this scope is required for storing mind maps as declared in the description, it also extends access to all files across the entire organisation, significantly expanding the potential attack surface.

Conversely, the \IncoherentRisky label is typically assigned only in the presence of obvious mismatches involving highly sensitive administrative scopes. For example, \textit{monday.com}, a project management platform, requests the \texttt{UserAuthenticationMethod.ReadWrite.All} permission, effectively allowing the app to change the authentication methods of all users in the organisation.

Interestingly, 13--15 applications ranked among the most anomalous are classified as \CoherentSafe. The motivations given by the LLM clarify that these cases rely mostly on resource-specific permissions. Although the choice of such permissions would be ideal from a security standpoint, the AD models flag these applications as anomalous because RSC permissions are rarely used.

Moving onto the least anomalous apps in Figure~\ref{fig:llm-result-heatmap-safe}, most of them are classified as \CoherentSafe, as expected. Only two applications, \textit{HubSpot} and \textit{Applauz}, are labelled as \IncoherentRisky. In both cases, the LLM highlights a clear semantic mismatch between the declared purpose and the requested permissions. These applications are described as notification-oriented tools, yet they request read access to Teams conversations. Notably, similar conversational read scopes appear with high frequency within the ``Notifications'' topic, thus not triggering the AD algorithms.

The high presence of \Insufficient classifications among the least anomalous cases highlights the anomaly detection models' struggle in detecting applications that are missing permissions required to accomplish declared functionalities.

We note that LLM judgements do not constitute ground truth. Rather, they provide an independent semantic signal indicating whether detected deviations correspond to meaningful differences in apparent access scope. The observed separation between anomalous and least-anomalous groups suggests that topic-aware anomaly detection 
is correlated with possible security and privacy risks for the users installing the applications.

{While topic analysis and anomaly detection do not provide a definitive or standalone method for assessing an application's risk profile, they can serve as valuable complementary tools for tenant administrators. By running the anomaly detection tool against an unknown application, the admin can obtain an additional metric to estimate the probability of the app showing potential misuse or malicious intent.}


\subsection{Manual inspection and verification}
\label{sec:results_evaluation_synthesis_of_findings}

\begin{table}
\caption{A sample of anomalous applications, the LLM verdict, and the most anomalous permissions they request, with the scores calculated as described in~\ref{sec:anom-permission}. Resource-Specific Consent (RSC) permissions are marked in bold. Permissions considered anomalous for not being requested are marked by a "!".}
\label{tab:manual-verification}
\centering
\begin{tabular}{p{2.8cm}p{2.6cm}p{2.4cm}p{4.5cm}c}
\toprule
\textbf{App Name} & \textbf{Topic} & \textbf{LLM Verdict} 
& \textbf{Top 5 Anomalous Permissions} & \textbf{Score} \\
\midrule

\multirow{5}{*}{Vacation Tracker}
& \multirow{5}{*}{Workspace Booking}
& \multirow{5}{*}{\textit{Coherent, Risky}}
& MailboxSettings.ReadWrite & 1.00 \\
& & & Group.Read.All & 0.83 \\
& & & MailboxSettings.Read & 0.66 \\
& & & \textbf{InAppPurchase.Allow.User} & 0.60 \\
& & & Team.ReadBasic.All & 0.60 \\

\midrule

\multirow{5}{*}{Seismic for Outlook}
& \multirow{5}{*}{File Management}
& \multirow{5}{*}{\textit{Coherent, Risky}}
& Group.Read.All & 0.86 \\
& & & Contacts.Read.Shared & 0.86 \\
& & & GroupMember.Read.All & 0.77 \\
& & & Directory.Read.All & 0.71 \\
& & & Contacts.Read & 0.66 \\

\midrule

\multirow{5}{*}{TeamOrgChart}
& \multirow{5}{*}{Intranet \& WebParts}
& \multirow{5}{*}{\textit{Insufficient}}
& email & 1.00 \\
& & & offline\_access & 1.00 \\
& & & openid & 1.00 \\
& & & profile & 1.00 \\
& & & User.Read & 0.33 \\

\midrule

\multirow{5}{*}{monday.com}
& \multirow{5}{*}{Project Management}
& \multirow{5}{*}{\textit{Incoherent, Risky}}
& User.Invite.All & 1.00 \\
& & & User.Export.All & 1.00 \\
& & & TeamsAppInstallation. ReadWriteSelfForUser.All & 1.00 \\
& & & User.ReadWrite & 1.00 \\
& & & UserAuthenticationMethod.Read.All & 1.00 \\

\midrule

\multirow{5}{*}{Officely}
& \multirow{5}{*}{Workspace Booking}
& \multirow{5}{*}{\textit{Incoherent, Risky}}
& ChatMember.ReadWrite.All & 1.00 \\
& & &\textbf{ChannelSettings.Read.Group} & 1.00 \\
& & & Channel.ReadBasic.All & 1.00 \\
& & & Chat.ManageDeletion.All & 1.00 \\
& & & TeamsAppInstallation. ReadWriteAndConsentSelfForTeam.All & 1.00 \\

\midrule

\multirow{5}{*}{Brochesia}
& \multirow{5}{*}{E-Learning}
& \multirow{5}{*}{\textit{Coherent, Risky}}
& \textbf{TeamsTab.Delete.Group} & 1.00 \\
& & & \textbf{TeamsTab.Create.Group} & 1.00 \\
& & & \textbf{Channel.Delete.Group} & 1.00 \\
& & & \textbf{Channel.Create.Group} & 1.00 \\
& & & \textbf{TeamsTab.Read.Group} & 0.83 \\

\midrule

\multirow{5}{*}{Vizerto}
& \multirow{5}{*}{AI Knowledge Base}
& \multirow{5}{*}{\textit{Coherent, Safe}}
& \textbf{ChatMessage.Send.Chat} & 1.00 \\
& & & \textbf{ChatMember.Read.Chat} & 0.82 \\
& & & \textbf{TeamMember.Read.Group} & 0.72 \\
& & & \textbf{Member.Read.Group} & 0.72 \\
& & & {! User.Read} & 0.12 \\

\midrule

\multirow{5}{*}{Podio Prod}
& \multirow{5}{*}{AI Knowledge Base}
& \multirow{5}{*}{\textit{Coherent, Risky}}
& Files.ReadWrite & 1.00 \\
& & & Contacts.Read & 0.82 \\
& & & Contacts.ReadWrite & 0.82 \\
& & & Mail.Read & 0.82 \\
& & & Files.Read & 0.72 \\

\bottomrule
\end{tabular}
\end{table}

To complement the LLM-assisted evaluation and assess its reliability, we conduct a blind manual inspection of a subset of applications. The goal is not to estimate detection precision statistically, but to qualitatively assess whether anomalous permission profiles correspond to semantically meaningful deviations in declared access scope. We select eight applications among those flagged as anomalous. Table~\ref{tab:manual-verification} summarises the apps' topic, verdict from the LLM, and the five most anomalous permissions, with their relative surprise score.

Overall, the manual inspection supports the effectiveness of our methodology. In six of the eight inspected cases, the anomalous permissions identified by our score correspond to privileges that appear excessive or difficult to justify given the application's declared functionality. The remaining cases highlight meaningful limitations of the approach, primarily due to the rarity of Resource-Specific Consent (RSC) permissions or to applications requesting fewer permissions than expected. Rather than serving as a definitive security verdict, topic-aware anomaly detection provides a practical mechanism to prioritise applications and permissions for further inspection, complementing existing security assessment processes.
In the following, we offer our findings, highlighting possible risks we found.

\textbf{Vacation Tracker.}
This app allows users to manage their paid time off within Microsoft Teams, Outlook, and Office.
It was correctly assigned to the ``Workspace Booking'' topic, given the planning nature suggested by its description. It requests 13 total permissions, some of which we judge as {excessive and not justified} by the features declared in the app's description. For instance, the \texttt{Calendars.ReadWrite} ($S=0.30$) was granted with Application scope, which, as introduced in Section~\ref{sec:microsoft_graph_permissions_m365}, allows the app to perform any operation to calendar events for \emph{every} user across the tenant. This permission is rarely requested by other apps in the same topic, thus raising the anomaly score. The app also requests the \texttt{MailboxSettings.ReadWrite} ($S=1.00$) permission, which also does not appear appropriate, as one would not expect a vacation tracker to modify a user's global mail settings. 

\textbf{Seismic for Outlook.}
Seismic is a sales management tool that facilitates content sharing and meeting workflows. This app adds Outlook integration to the Seismic software. It is correctly assigned to the ``File Management'' topic, given its focus on file sharing. Of the 10 total permissions requested by the app, we notice \texttt{Directory.Read.All} ($S=0.71$), an Application-scoped permission which grants the ability to read the \emph{entire} organisation's directory structure. This level of access does not seem justified based on its stated functionality. In fact, this permission is not requested by other apps (at large).

\textbf{TeamOrgChart.}
This app is designed to generate and manage organisational charts and is correctly assigned to ``Intranet \& WebParts'', which focuses on SharePoint Web parts and related tools. This app requests only 5 minimal, delegated sign-in permissions. Of them, the only major permission is \texttt{User.Read} ($S=0.33$), allowing the app to read the information of users who provide consent. We consider this as an example of \emph{under}-privileged application that our models flagged as anomalous because the number and type of requested permissions differs from those of other apps in the same topic. The LLM correctly flag this app permissions as insufficient because it requests fewer permissions than the minimal necessary to perform the required functionality. This may also raise concerns about the app actual purpose.


\textbf{Monday.com.}
The application is designed to simplify the management of a team projects, providing a collaboration platform for organising and tracking tasks, team members, and custom workflows. Its domain fits well with the assigned topic ``Project Management''. Among the 14 requested permissions, we find \texttt{UserAuthenticationMethod.ReadWrite.All} ($S=1.00$), effectively requesting to read and modify the authentication methods of all users in the organisation. Authentication methods include things like a user’s phone numbers and Authenticator app settings, opening the door for possible abuse and data leaks. From the app's description, the need for such permissions is questionable (and no other applications in this category require such high privileges).

\textbf{Officely.}
The application is a desk and workspace booking tool integrated within Microsoft Teams. It was correctly assigned to the topic ``Workspace Booking''. Despite some of the 14 requested permissions being justified (e.g., \texttt{User.Read.All} ($S=0.32$) to identify which users are booking desks), the overall permission set is excessive considering the application's purpose stated in its description. Permissions like \texttt{ChatMember.ReadWrite.All} ($S=1.00$) and \texttt{Chat.ManageDeletion.All} ($S=1.00$) are questionable and pose a security risk, as they allow the application to add or remove members from any chat and delete any chat message across the entire organisation.

\textbf{Brochesia.}
This application is designed to help on-site workers solve technical problems via Teams using Augmented Reality. We consider its classified topic (E-Learning) appropriate, even if not exactly a classic e-learning platform. All the 8 permissions requested by it are resource-specific (RSC), therefore restricting management to the specific Team it is installed into. Although the choice of using RSC permission would be ideal from a security standpoint, the AD flags this application as anomalous, as RSC permissions are rare within our dataset. This is a limitation of our methodology, which focuses more on the frequency of permissions rather than their adherence to security best practices. 

\textbf{Vizerto.}
This app declares itself as a GenAI assistant to provide context-specific answers to business- and product-related questions within an organisation. It was correctly assigned to the topic ``AI Knowledge Base'', in line with its description. The app requests 4 RSC permissions only. These permissions are limited to reading specific chat, group and team member information (e.g., \texttt{TeamMember.Read}, $S=0.72$). It also includes the \texttt{ChatMessage.Send.Chat} ($S=1.00$) permission, which allows the assistant to send messages in user chats. We consider these permissions appropriate based on the app's description. This app likely shares the same reason as Brochure for flagging, given its heavy reliance on RSC permissions.

\textbf{Podio.}
The application provides no-code tools for building software solutions and setting up custom workflows.
It was assigned to the topic ``AI Knowledge Base'', although it could also fit the ``Project Management'' topic. Among the 9 permissions it requests, we notice some of them are quite high-level, such as \texttt{Files.ReadWrite} ($S=1.00$) and \texttt{Contacts.ReadWrite} ($S=0.82$). The LLM mark thus the app as risky. Nonetheless, these permissions appear well aligned with the app’s intended functionality: its workflow automation goals inherently require access to a wide range of features to support organisational processes, which could be seen as justified. The permissions are thus coherent for the LLM.

%% file: text/7-related-work.tex
\section{Related works}
\label{sec:related}

The security and privacy of application ecosystems have been widely investigated across different domains. Numerous works have analysed permission misuse, over-privileged access and data collection practices in Google's Play Store for Android \cite{yang2025guidelines, g-cata_2024, BERTDetect_2025}, Apple's App Store for iOS \cite{mohd2024ios, scoccia2022empirical}, and the general Web ecosystem \cite{fernandez2025permissions}. More recently, similar concerns have emerged for large language model (LLM) ecosystems where large-scale analyses have uncovered misleading descriptions, abusive functionality, and excessive data collection among GPT-based chat systems \cite{hou2025security, wu2025depth}.

Few works exist involving the detection of suspicious consent grants in the Microsoft ecosystem, i.e., risks of users granting access to their data as a result of a phishing or unauthorised code execution attack. Previous works in the literature focus on attacking the ecosystem via machine-learning-based approaches~\cite{zannone2023security} or exploiting security vulnerabilities~\cite{chen_2022_experimental}.
Compared with these works, our study considers permissions that users voluntarily grant to applications that do not try to evade system defences. Another work~\cite{potcoveanu2025analyzing} looks at network traces to detect potential information leaks from applications installed from the Microsoft Store. However, transmission of such data is unaware of the application's behavioural context, which makes it hard to detect possible violations.

Researchers have started exploring the possibility of using easily collectable metadata features to build stronger context and improve detection. This is enforced by the fact that anomalous behaviour always depends on the context, and this is particularly true in the field of application security. For instance, a feature that is anomalous for one application can be perfectly normal for another \cite{CHABADA_Gorla_2014}. Late improvements in Natural Language Models and the introduction of Neural Topic Modelling (NTM) Techniques, such as Latent Dirichlet Allocation (LDA) \cite{LDA_Blei_2003} and BERTopic \cite{BERTopic_Grootendorst_2022}, have inspired researchers to extract semantic features to intersect context and behaviour. 

A first attempt to use NTM techniques in malware detection was made by Gorla et al. \cite{CHABADA_Gorla_2014} who presented CHABADA, a system that was able to flag novel malware by leveraging NTM and clustering (K-Means in their case) to extract topics from the application descriptions.
Further improvements were made by Ranaweera et al. who proposed BERTDetect~\cite{BERTDetect_2025}, a system that leverages BERTopic to extract topics, achieving improvements in False Negative Rate (FNR) and True Positive Rate (TPR). In our work, we build on similar techniques to identify homogeneous apps classes and then run AD algorithms.

Considering NTM, BERTopic~\cite{BERTopic_Grootendorst_2022} is one of the most recent and effective models for topic extraction, as it leverages the power of transformer-based embeddings, unlike probabilistic bag-of-words models such as LDA \cite{chamomile_topic_modeling_2025} or TF-IDF. BERTopic is proven to be effective in capturing semantic similarity among documents, remaining competitive across a wide range of benchmarks involving traditional models in topic modelling. 
For anomaly detection, we rely on well-established algorithms. Recent approaches have explored the use of deep neural network–based models, such as autoencoders, for anomaly detection~\cite{pang2021deep}. However, these methods typically require large amounts of training data and introduce additional computational and modelling complexity, which are not well-suited to our limited dataset and purely unsupervised setting.

%% file: text/8-conclusion.tex
\section{Conclusions}
\label{sec:conclusion}

We presented the first systematic measurement of the Microsoft 365 (M365) third-party application ecosystem from a permission-centric security perspective. By combining public APIs, automated tenant-side deployment, and operational tenant data, we crawled more than 8,000 applications, of which only 1,069 expose both textual descriptions and OAuth permission sets. This limited coverage itself highlights the lack of transparency and the fragmented nature of the current M365 application ecosystem.

To assess whether applications request permissions consistent with their declared functionality, we proposed a topic-aware analysis pipeline combining Neural Topic Modelling with unsupervised anomaly detection. Rather than comparing applications across the entire ecosystem, our approach models permission profiles relative to semantically similar peers, enabling the identification of both over-privileged and under-privileged applications. Using a conservative consensus criterion, we identified 139 anomalous applications (13\% of the analysed dataset).

LLM-assisted evaluation and blind manual inspection support the effectiveness of the proposed methodology, showing that many detected anomalies correspond to permission requests that appear difficult to justify given the application's declared functionality. Although metadata alone cannot prove misuse or malicious intent, topic-aware anomaly detection provides a useful signal for prioritising applications for further security review.

Overall, our findings expose structural opacity and non-uniform privilege allocation in the M365 ecosystem, underscoring the need for improved transparency and automated permission auditing in enterprise environments. We believe the proposed pipeline can provide tenant administrators with a practical decision-support tool, helping them identify applications whose requested permissions deviate from those of functionally similar peers before granting consent, thereby supporting more informed decisions and promoting the principle of least privilege.

%% file: text/998-appendix-short.tex
\section{Ethics}
The ethical considerations about this work mostly reside in the data collection phase. First, we took all the precautions to prevent the crawling of applications' permissions via the Microsoft Marketplace from negatively impacting the destination platform: we only performed a few thousand operations on the platform in a range of more than 48 hours, a negligible volume for a company with resources such as Microsoft's. Second, while collecting app permissions from the University tenant, we received from the system administrators \emph{anonymised} data about installed applications, which did not contain any identifier that could be linked to any University student or faculty.

\section{BERTopic parameter selection}
\label{sec:bertopic-parameter-selection}

\begin{table}
    \centering
    \caption{Final selected BERTopic configurations.}
    \begin{tabular}{ll}
        \toprule
        \textbf{Parameter} & \textbf{Base Value}\\
        \midrule
        SBERT\_model & all-MiniLM-L6-v2 \\
        n\_components & 35 \\
        n\_neighbors & 5 \\
        min\_cluster\_size & 10 \\
        min\_samples & 10 \\
        min\_df & 5 \\
        max\_df & 0.95 \\
        ngram\_range & (1, 2) \\
        top\_n\_words & 10 \\
        \bottomrule
    \end{tabular}
    \label{tab:bertopic_parameters}
\end{table}

As described in Section \ref{sec:topic-retrieval}, we perform a ``bottom-up'' hyper-parameter tuning process selecting the set of parameters which can have the highest impact on the model performance. We proceed with the tuning process, starting from one model and running each configuration of parameters with three different random states. At each step, we select the best configurations to include in the next search for the upper model and its parameters, and so on. For the sake of clarity, we leave the details to a separate document we will link to upon acceptance, to preserve the authors' anonymity. We simply report the BERTopic final parameters in Table~\ref{tab:bertopic_parameters}.

\section{Extracted topics}
\label{sec:extracted-topics}

The list of extracted topics, their meaning, and the number of associated applications is detailed in Table~\ref{tab:extracted_topics}.

\begin{table}
\caption{Representative words, interpretation, and application count for each topic.}
\centering
\resizebox{\columnwidth}{!}{
\begin{tabular}{cllc}
\toprule
\textbf{ID} & \textbf{Top-5 Terms} & \textbf{Interpretation} & \textbf{Count} \\
\midrule
0  & (crm, file, template, attachment, send) & File Management & 122 \\
1  & (360, student, learning, course, training) & E-Learning & 65  \\
2  & (ticket, notification, onboarde, core, incident) & Notifications & 65  \\
3  & (office, room, book, desk, space) & Workspace Booking & 59  \\
4  & (asset, security, permission, software, expense) & Security & 75  \\
5  & (calendar, event, meeting, appointment, scheduling) & Scheduling \& Calendar & 63  \\
6  & (idea, board, meeting, visual, brainstorm) & Collaborative Brainstorming & 62  \\
7  & (sharepoint, page, card, viva, connection) & SharePoint Integration & 49  \\
8  & (birthday, web, sharepoint, webpart, navigation) & Intranet \& WebParts & 45  \\
9  & (signature, sign, document, signing, digital) & Digital Signatures & 33  \\
10 & (task, project, checklist, board, gantt) & Project Management & 50  \\
11 & (sms, call, communication, message, agent) & Communication Tools & 31  \\
12 & (recognition, culture, employee, recognize, peer) & Employee Recognition & 29  \\
13 & (ai, knowledge, answer, enterprise, unleash) & AI Knowledge Base & 50  \\
14 & (meeting, ai, summary, transcript, meet) & AI Meeting Assistants & 37  \\
15 & (survey, video, feedback, quiz, meeting) & Surveys \& Feedback & 28  \\
16 & (break, copilot, habit, journey, movement) & Well-being \& Productivity & 36  \\
17 & (language, translation, meeting, multilingual, ai) & AI Translation & 19  \\
18 & (confluence, tab, view, analytic, meeting) & Confluence Integration & 55  \\
19 & (form, approval, travel, sharepoint, component) & Forms \& Workflows Design & 25  \\
20 & (pdf, pdfs, document, convert, file) & PDF Management & 15  \\
21 & (directory, employee, people, hr, profile) & Human Resources & 20  \\
22 & (expense, project, task, budget, claim) & Expense Management & 24  \\
23 & (legal, ai, firm, matter, capture) & Legal AI Assistant & 12 \\ 
\bottomrule
\end{tabular}
}
\label{tab:extracted_topics}
\end{table}



\section{Anomaly detection model tuning}
\label{sec:anomaly-detection-model-tuning}

\subsection{Injecting Outliers}
\label{subsec:results_building_injected_outliers}
As mentioned in Section~\ref{sec:ad-method}, we used three distinct anomaly detection algorithms to identify outliers in each topic: One-Class SVM (OC-SVM), Isolation Forest (iForest), and Local Outlier Factor (LOF). This means that for each algorithm we train and evaluate a separate model for each topic. Before applying the models directly on the topics, we first look to assess their performance on identifying anomalies and select the best hyper-parameters for each algorithm. The complete list of parameters for each level of difficulty are reported in Table \ref{tab:injected_outliers_parameters_synthetic}.

\begin{table}
    \centering
    \caption[Synthetic difficulty levels parameters.]{Parameters used for synthetic outlier injection under different perturbation difficulty levels.}
    \begin{tabularx}{\columnwidth}{lXc}
        \toprule
        \textbf{Difficulty} & \textbf{Parameter} & \textbf{Value} \\
        \midrule
        \multirow{2}{*}{Easy}   & \texttt{num\_common\_to\_remove} & 4 \\
                                & \texttt{num\_rare\_to\_add} & 4 \\ 
        \midrule
        \multirow{2}{*}{Medium} & \texttt{num\_common\_to\_remove} & 2 \\
                                & \texttt{num\_rare\_to\_add} & 2 \\
        \midrule
        \multirow{2}{*}{Hard}   & \texttt{num\_common\_to\_remove} & 1 \\
                                & \texttt{num\_rare\_to\_add} & 1 \\
        \bottomrule
    \end{tabularx}
    \label{tab:injected_outliers_parameters_synthetic}
\end{table}

We evaluated the performance of each model on each injection strategy (e.g., Hard Synthetic) plotting the average ROC curve across all topics. On the x-axis of the ROC curve we report the False Positive Rate (FPR), which represents the percentage of native outliers detected by the model. On the y-axis we report the True Positive Rate (TPR), representing the percentage of injected outliers correctly flagged by the model. The objective is to maximize the TPR while keeping an acceptable FPR (i.e., not flagging too many native applications as outliers).

\begin{figure}
    \centering


    \begin{subfigure}{.45\columnwidth}
        \centering
        \includegraphics[width=\columnwidth]{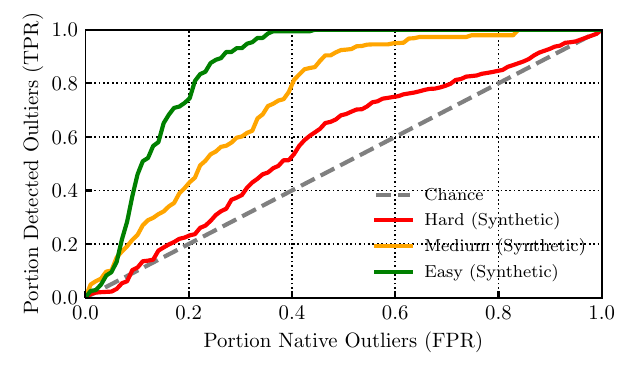}
        \caption{Average ROC curves for LOF.}
        \label{fig:lof_roc_curves_injected_outliers_synthetic}
    \end{subfigure}
    \hspace{.05\columnwidth}
    \begin{subfigure}{.45\columnwidth}
        \centering
        \includegraphics[width=\columnwidth]{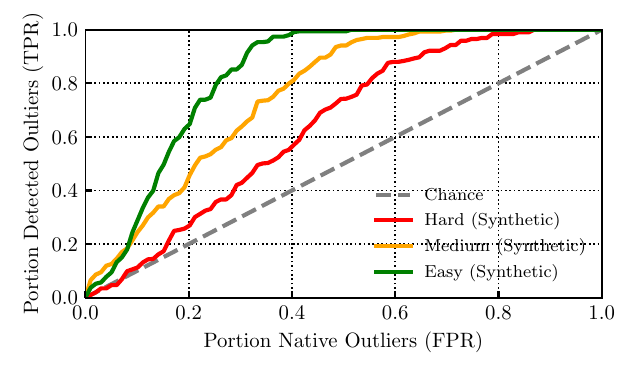}
        \caption{Average ROC curves for iForest.}
        \label{fig:isolation_forest_roc_curves_injected_outliers_synthetic}
    \end{subfigure}

    \caption[Model Robustness Analysis]{Average ROC curves for OC-SVM, LOF, and Isolation Forest with synthetic outlier injection across all topics.}
    \label{fig:all_models_roc_curves}
\end{figure}

As shown in the average ROC curves in Figure \ref{fig:all_models_roc_curves}, the OC-SVM reflects the difficulty levels almost perfectly for the easy level. 
As expected, the medium and hard levels result more challenging for the model.

Both LOF and iForest models show more difficulty in solving the Synthetic strategy compared to OC-SVM, although they still display decent performance.


The difficulty levels chosen for each algorithm are reported in Table \ref{tab:selected_injected_outliers_difficulty_levels} along with the best global hyperparameters identified after the grid search with k-fold cross-validation.

\begin{table}
    \centering
    \caption[Selected Difficulty Levels for Anomaly Detection Algorithms]{Selected difficulty levels and hyperparameters for each anomaly detection algorithm.}
    \begin{tabular}{lcc}
        \toprule
        \textbf{Algorithm} & \textbf{Difficulty Level} & \textbf{Best Hyperparameters} \\
        \midrule
        OC-SVM         & Medium & \texttt{nu}=0.1 \\
        LOF           & Medium   & \texttt{k}=5 \\
        iForest & Medium   & \texttt{n\_estimators}=100 \\
        \bottomrule
    \end{tabular}
    \label{tab:selected_injected_outliers_difficulty_levels}
\end{table}

\subsection{Model Performance on Injected Outliers}
\label{subsec:results_model_performance_injected_outliers}
Once selected the best hyper-parameters for each algorithm, we evaluate the performance of each model on the injected outliers. 
All anomaly detection algorithms used in this project are designed to work with thresholds and anomaly scores. Specifically, OC-SVM has a threshold which is implicitly set to zero, as the anomaly score is the distance from the separating hyperplane, while LOF and iForest require the setting of a threshold on the anomaly scores to classify an application as an outlier. For these two algorithms, we also provide an analysis of the optimal thresholds for each topic, selecting the best global threshold across all topics based on the median of optimal thresholds. The optimal threshold for each topic is identified by maximizing the Youden's Index, i.e., $TPR - FPR$.
The summary of the selected thresholds and the corresponding performance (TPR, FPR) is reported in Table \ref{tab:thresholds_performance}.

\begin{table}
    \centering
    \caption[Summary of Selected Thresholds and Performance]{Summary of selected thresholds and corresponding performance for each anomaly detection algorithm on injected outliers.}
    \begin{tabular}{lccc}
        \toprule
        \textbf{Algorithm} & \textbf{Selected Threshold} & \textbf{TPR} & \textbf{FPR} \\
        \midrule
        OC-SVM         & 0 (default)      & 0.91 & 0.32 \\
        LOF           & -1.29 (median)   & 0.61 & 0.31 \\
        Isolation Forest & 0.39 (median)   & 0.79 & 0.39 \\
        \bottomrule
    \end{tabular}
    \label{tab:thresholds_performance}
\end{table}